**Title: Visualizing Symmetry Broken Chern Insulators and their Quantum Melting**

**Authors:** Minhao He[1,*], Yen-Chen Tsui[1,*], Ran Peng[2], Kenji Watanabe[3], Takashi Taniguchi[4], Oskar Vafek[2,5], Ali Yazdani[1,†]

**Affiliations:**
[1] *Joseph Henry Laboratories and Department of Physics, Princeton University, Princeton, NJ 08544, USA*
[2] *Department of Physics, Florida State University, Tallahassee, Florida 32306, USA*
[3] *Research Center for Functional Materials, National Institute for Materials Science, 1-1 Namiki, Tsukuba 305-0044, Japan*
[4] *International Center for Materials Nanoarchitectonics, National Institute for Materials Science, 1-1 Namiki, Tsukuba 305-0044, Japan*
[5] *W. I. Fine Theoretical Physics Institute and School of Physics and Astronomy, University of Minnesota, Minneapolis, Minnesota 55455, USA*

* These authors contributed equally to this work.
† Corresponding author email: yazdani@princeton.edu

**Abstract:** In the presence of a magnetic field, electronic states of moiré quantum materials develop a Hofstadter spectrum that provides a unique setting for studying the interplay between band topology and strong electron-electron interaction. Using scanning tunneling microscopy, we study Hofstadter's states in bilayer graphene aligned with hexagonal BN and directly visualize the formation of interaction-driven symmetry breaking Chern insulators. Our measurements reveal the formation of phases that double, triple or quadruple the moiré unit cell at fractional filling of the Hofstadter bands, as well as states with complex intra-unit-cell wavefunctions. We visualize two distinct quantum phenomena in different Chern states, including quantum melting driven by the appearance and proliferation of topological defects, and a quantum transition co-occurring with phase competition and separation.

**Main Text:**
The interplay between magnetic fields and lattice periodicity gives rise to fractal energy bands, as first predicted by Hofstadter(*1*). Originally proposed in electronic systems, the Hofstadter's model has since been realized in a wide range of experimental platforms, such as waveguide array(*2*), optical lattice(*3*, *4*), superconducting qubit chain(*5*), or GaAs/AlGaAs heterostructure by fabricating lateral artificial superlattice(*6*). Moiré superlattices in stacked van der Waals materials, formed by twist or lattice mismatch, provide an ideal platform for studying the Hofstadter's spectrum and its topology. Early experiments(*7–9*) have demonstrated that topological states, called Chern insulators (CI), emerge when the Hofstadter subbands are filled up to a single particle gap. Each CI can be described by two quantum indices: the Chern number(*10*) and the band filling index *s*(*11*). The fractal Hofstadter energy spectrum giving rise to these incompressible states has recently been detected using scanning tunneling spectroscopy (STS)(*12*).

Unusual correlated and topological phases(*13*, *14*) emerge at partial fillings of the Hofstadter subbands when their bandwidth becomes narrow and comparable with Coulomb interaction. In this situation topological invariants ($C$ and/or *s*) alone do not offer a complete description, as they are insensitive to presence of any broken symmetries. As a result, fundamentally distinct many-body states can share identical topological indices while differing in their underlying microscopic symmetry structures. Theoretical predictions of strongly interacting Hofstadter's states include two possible ground states: incompressible states with quasi-particles associated with fractional charges(*15–18*), and phases that break the moiré translational symmetry(*19*). Despite their presence in various graphene-based moiré materials(*13*, *14*, *20–23*), there has been no microscopic study of the wavefunctions and symmetries of these strongly interacting Hofstadter states(*24*). An equally important open question is the nature of the quantum phase transitions that drive the loss of broken symmetry or topological order of such phases.

Recent scanning tunneling microscopy (STM) measurements have demonstrated the capability of imaging delicate electronic phases with minimal tip-sample work function mismatch(*25–27*). Here we report direct imaging of the symmetry broken Hofstadter's states and visualize their associated quantum melting in Bernal-stacked bilayer graphene (BLG) fully aligned with h-BN using STM. We use spectroscopic measurements to identify various CI insulating phases (by measuring their $C$ and *s*), and find symmetry broken Chern insulators (SBCI) that either double, triple or quadruple the moiré unit cell, some with intricate intra-unit-cell structures. Our approach can be applied to also directly visualize the process of quantum melting of SBCIs and CIs that has remained inaccessible to macroscopically averaged techniques.

**Moiré superlattice and the emergence of Hofstadter states**
The BLG devices studied here are fully aligned with the h-BN substrate. Figure 1A shows a typical constant current topographic image within an ultraclean area, with an extracted moiré periodicity $\lambda$ of 14.28 nm, consistent with superlattice periodicity of a fully aligned BLG/h-BN(*28*, *29*). A flux ratio as large as $\phi/\phi_o \approx 0.6$ is reached at $B$ = 14 T, enabling real-space imaging of electronic wavefunctions in the fractal Hofstadter's spectrum. Here $\phi = B(\sqrt{3}/2\,\lambda^2)$ is the magnetic flux through one moiré unit cell, and $\phi_0 = h/e$ is magnetic flux quantum, where $h$ is Planck's constant and $e$ is the elementary charge).

In Fig. 1B, we show STS as a function of gate voltage $V_G$ measured in this region at $B$ = 13.8 T. Similar to previous work on misaligned BLG/hBN samples(*26*) (fig. S1, section S1 (*30*)), a strong intrinsic displacement field ($D_0$) in our devices (hBN on only one side of BLG) lifts the layer degeneracy of the eight lowest Landau levels (LLLs) and favors filling the bottom layer polarized (hBN proximate) LLs first. Being further from the tunneling tip, this makes the spectral weight at $V_G < 0$ weaker than $V_G > 0$. Nevertheless, we observe significant differences from the hBN misaligned samples, including a series of additional incompressible states (Fig. 1B). When the electrons are polarized to the bottom layer ($V_G < 0$), we observe sharp peak energy shifts (also visible in remote bands) signifying chemical potential jumps across incompressible states at carrier densities distinct from the quantum Hall ferromagnets (QHFMs) at LL filling factor $\nu_{LL}$ = -3, -2, -1 (Fig. 1B, fig.S2, section S2 (*30*)). These states highlight the strong moiré effect that arises when electrons are polarized close to the moiré interface, and can be assigned to CIs in the Hofstadter spectrum. This is further supported by the CIs are only observed above $B$ = 6 T (fig. S3, section S3 (*30*)). When the electrons are polarized to the top layer ($V_G > 0$), we still observe the additional Hofstadter's states, but with much weaker spectral features (fig. S4, section S4 (*30*)), indicating a less pronounced moiré effect.

**Identifying Chern insulators**

We first identify each CI by determining their quantum indices: the Chern number $C$ and the band filling index $s$. We closely inspect the Hofstadter states at $V_G < 0$, focusing on the spectral features near the Fermi level ($V_B = 0$). The incompressible gaps are identified as peaks in the tunneling threshold gap $\Delta_t$ defined with an onset tunneling current (fig. S5, section S5 (*30*)). We perform magnetic field dependent STS to construct the Wannier diagram (Fig. 1C), a plot of carrier density $n$ versus magnetic flux per moiré unit cell $\phi/\phi_o$ in which each incompressible state follows a straight line (fig. S6 and S7). From the slope and intercept of these lines, we determine the Chern number $C$ and band filling index $s$ using the Diophantine relation $n/n_s = C\,\phi/\phi_0 + s$. Here $n_s = 1/(\sqrt{3}/2\,\lambda^2)$ is the carrier density corresponding to one electron per moiré unit cell with periodicity $\lambda$. Our measurements span a large range of magnetic fields, importantly in the vicinity of magnetic flux ratio $\phi/\phi_0$ = 1/2, where we observe spectral features of a dispersive Hofstadter subbands as predicted (fig. S8, section S6 (*30*)). We classify the observed states (fig. S9, section S7 (*30*)) and summarize a full phase diagram in Fig. 1C, including similar measurements performed at $V_G > 0$ (fig. S10 and S11).

The moiré potential impacts mostly the phase diagram at $V_G < 0$, where the carrier density is tuned in the layer closest to the moiré potential. In this regime, we observed four sets of Chern insulators (CIs) that follow the sequence $(C, s) = (-2(N + s_0), N)$ $(N = 0, 1, \dots)$, with each set crossing $\phi/\phi_o$ = 1/2 at a different density $-s_0$ = 0, -1, -1.5, and -2. These CIs can be understood as arising from the Landau quantization of the maxima or minima of dispersive magnetic subbands at $\phi/\phi_o$ = 1/2 (Fig. 1D and E). In other words, when the flux $\phi$ is shifted from half flux, electron orbits near the extrema of these dispersive magnetic subbands become quantized into LLs according to the detuned flux $\delta\phi = \phi/\phi_o - 1/2$. The magnetic subbands at $\phi/\phi_o$ =1/2 can be constructed using the eigenfunctions of magnetic translation group(*31*) (MTG). In general, for flux $\phi/\phi_o = p/q$ with Landau gauge $\mathbf{A} = (0, Bx, 0)$, the magnetic lattice momentum is written as $\mathbf{k} = k_1\mathbf{g}_1 + k_2\mathbf{g}_2$ where $k_1 \in [0,1)$, $k_2 \in [0, 1/q)$. The discrete magnetic translation symmetry of the moiré superlattice requires the energy of subbands to be degenerate under $k_1 \to k_1 + 1/q$ [section S8 (*30*)]. Therefore, at flux $\phi/\phi_o$ =1/2, the magnetic subbands and the LLs that emerge when flux is shifted away from half flux, are two-fold degenerate as long as the translation symmetry is preserved(*31*) (Fig. 1D). This in turn explains why the Chern numbers of translation symmetry preserving states in each set differ from each other by even integers. To illustrate this point, consider depleting two Landau levels emanating from the maxima of the charge neutrality point valance subbands at half flux, as in the schematic in Fig 1E. The filling of such a state is $n/n_s = -2\delta\phi = -2(\phi/\phi_o - 1/2)$, corresponding to $(C, s) = (-2, 1)$ (Fig. 1E). More generally, depleting $2N$ such LLs gives the sequence $(-2N, N)$. The sequences with nonzero $s_0$ can be similarly understood as first placing the Fermi level inside the Chern $-2s_0$ gap between the dispersive subbands at half flux and then depleting $2N$ half flux shifted LLs. The filling of such a state is then $n/n_s = -2s_0\phi/\phi_o - 2N(\phi/\phi_o - 1/2)$ resulting in observed sequence $(C, s) = (-2(N + s_0), N)$. Note that because each subband carries an integer Chern number, $2s_0$ is an integer (which can be even or odd).

More importantly, we also observe interacting Hofstadter states with integer *C* but fractional *s*(*13*, *14*). Plausible ground state candidates for these states include charge density waves with broken moiré translation symmetry(*14*, *21–23*). With translation symmetry broken in one direction, the degeneracy between the states at $k_1$ and $k_1 + 1/2$ is lifted, which allows a single half flux shifted LL to be filled atop Chern -4 gap, resulting in (-3,-1/2), or to be emptied starting from the charge neutrality point Chern 0 gap, resulting in (-1,1/2) (Fig. 1F). However, such broken symmetry states have never been experimentally established since previous studies lack microscopic information.

**Visualizing Chern insulators and their broken symmetries**
To determine whether the Chern insulators at fractional filling break any translational symmetry, we use STM to directly visualize the spatial structure of their electronic density of states. We measured the spatial modulation of the tunneling current $\delta|I_{dc}|$ at small bias voltages [section S9 (*30*)]. As expected, the integrated density of state maps for the CI with (*C*, *s*) = (-2, 1) preserves the moiré translational symmetry as shown in Fig. 2A. This is further confirmed in their structure factor $S(\vec{q})$ shown in Fig. 2F [section S9 (*30*)], with six sharp moiré Bragg peaks (in red hexagons) in $C_3$ symmetry. Notably, a CI with (*C*, *s*) = (-4, 1), while preserving moiré translational symmetry, develop an emergent Kagome charge order with intra moiré unit cell structure (Fig. 2B). This feature gives rise to the higher order Bragg peaks (blue circles) in Fig. 2G. The rich intra moiré unit cell structure highlight importance of coulomb interaction within a unit cell(*32*), even in the cases of CIs with integer (*C*, *s*).

Signature of breaking the moiré translational symmetry appears in the STM images of CIs with fractional *s*, such as the state (*C*, *s*) = (-1, 1/2) shown in Fig. 2C. Instead of the original triangular moiré translation symmetry, we measure a stripe-like charge order with reconstructed unit cell (green rectangles). The structure factor $S(\vec{q})$ maps (Fig. 2H) constructed from these images show the absence of $C_3$ symmetric moiré Bragg peaks (red hexagons), and the emergence of new $C_2$ symmetric Bragg peaks (green squares). The dominant Bragg peaks appear at half of the reciprocal vector $\boldsymbol{b}$ (denoted in Fig. 2F) of the moiré superlattice, showing that the real-space periodicity of the stripe charge order is twice the moiré lattice constant. Our observation is consistent with a charge density wave (CDW) doubling the moiré unit cell at half filling of the Hofstadter's subband. Our observation of CDWs with a finite Chern number *C* is distinct from previous reports of moiré commensurate CDWs, termed generalized Wigner crystals(*33*), which are topologically trivial.

Our imaging experiments establish the ground state for the case of integer $C$ and fractional $s$ to be the anticipated symmetry broken Chern insulators (SBCIs). Our experiments demonstrate this behavior for all incompressible states with integer $C$ and fractional $s$, including those at $V_G > 0$ (Fig. 2, D and E, I and J, fig. S13 to S15, section S10 (*30*)). Figure 2D shows STM imaging of ($C$, $s$) = (2, 1/3) state, which shows three times the size of the moiré unit cell (orange parallelogram). The corresponding $S(\vec{q})$ map (Fig. 2I) also shows the corresponding $C_3$ symmetric Bragg peaks (orange hexagons), appearing at the reciprocal vectors of $|\boldsymbol{b}|/\sqrt{3}$. This is consistent with the simplest picture of one-third filling of a Hofstadter subband, in which one electron charge is distributed over three moiré unit cells. In addition, we show that the broken translation symmetry can take on forms more complicated than existing predictions, as in the case of ($C$, $s$) = (1, 1/2) (Fig. 2E). The band filling index $s$ indicates a half-filled Hofstadter subband, which in the simplest case corresponds to one charge in two moiré unit cells. Instead, we observe electronic structure that quadruples the unit cell (pink parallelogram). Correspondingly, six sharp peaks with magnitude of $|\boldsymbol{b}|/2$ appear in the $S(\vec{q})$ map (Fig. 2J). Our real-space imaging suggests the ($C$, $s$) = (1, 1/2) SBCI corresponds to a quadrupled reconstructed unit cell, which incorporates two electron charges.

In addition to the rich inter-unit-cell reconstruction, we observe complicated intra-unit-cell structures of these SBCIs, that are either due to higher harmonics of the moiré potential, or formed due to the strong electron-electron interactions. Moreover, we also find the orientation of the (-1, 1/2) stripe SBCI in the same field of view rotates upon slight charge doping (fig. S16, section S11 (*30*)), suggesting that this electronic order is not strongly coupled to the underlying lattice or sample strain. However, whether SBCI forms due to electron-phonon coupling, or a pure electronic Coulomb interaction remains an open question.

**Quantum melting and topological defects**

Our imaging technique also enables a direct study of the process of quantum melting of SBCI tuned by electron density, as well as determining the role of topological defects that may be present as these phases approaching melting. Imaging the stripe SBCI ($C$, $s$) = (-1, 1/2) (Fig. 3A) over a large field of view, we observe a number of notable features, such as domains with stripes in different orientation, and also boundaries that separate them. The three stripes give rise to the reciprocal wavevector $\boldsymbol{q}_{i=1,2,3}$ shown in the FFT map (Fig. 3B), at half of the original moiré wavevector $\boldsymbol{b}_i/2$. We characterize the properties of this intertwined SBCI using an analysis that separately investigates the individual stripe component, following methods previously used to analyze similar phenomena in stripe phases(*34*, *35*) [section S12 in (*30*)].

Figures 3, D to F, show the relative phase $\varphi_{\boldsymbol{q}_1}(\boldsymbol{r})$, the SBCI amplitude distribution $\rho^{o}_{\boldsymbol{q}_1}(\boldsymbol{r})$, and the cosine charge modulation $\cos(\boldsymbol{q}_1 \cdot \boldsymbol{r} + \varphi_{\boldsymbol{q}_1}(\boldsymbol{r}))$ of the stripe SBCI associated with $\boldsymbol{q}_1$, respectively. The same operation can be repeated for $\boldsymbol{q}_2$ and $\boldsymbol{q}_3$ (fig. S17). Figure 3C shows the normalized SBCI amplitude distribution of the three wavevectors $\boldsymbol{q}_i$, represented by red, green and blue respectively. We found the amplitude distribution analysis shows that all three directions can coexist with each other.

Focusing on the relative phase $\varphi_{\boldsymbol{q}_1}(\boldsymbol{r})$ shown in Fig. 3D, we discuss the presence of topological defects within a single stripe orientation. The two topological defects are dislocations and anti-dislocations, corresponding respectively to where stripes terminate and where they merge. In the phase map, they appear as singular points: a clockwise loop around a dislocation gives a $+2\pi$ phase winding (black), while a clockwise loop around an anti-dislocation gives a $-2\pi$ phase winding (red). We observe a pair of dislocation and anti-dislocation; they are responsible for the sharp boundary that separates two patches of the stripes shown in Fig. 3F, with an offset of one moiré superlattice constant. An interesting question is whether a (anti-)dislocation carries fractional charge.

The topological defects are not pinned to specific topographic features, such as lattice point defects. Instead, we observe that the topological defects change locations when doping the SBCI (Fig. 3H). We show a zoomed-in STS at gate voltages $V_G$ nearby the ($C$, $s$) = (-1, 1/2) SBCI in Fig. 3G. We image the charge order with broken translation symmetry at eight different $V_G$ (fig. S18 and S19) and analyze the locations and types of topological defects as in Fig. 3H. The total number of topological defects, summing all three wavevectors $\boldsymbol{q}_i$, are plotted in Fig. 3I as a function of $V_G$. We observe a trend of increasing the total number of (anti-)dislocations as the system is doped away from the center of the incompressible gap near $V_G$ = -0.168 V. Our measurements indicate that the quantum phase transition for this SBCI to a compressible phase is mediated by proliferation of topological defects, consistent with a Kosterlitz-Thouless (KT) transition(*36*, *37*).

**Quantum melting and phase separation**

Finally, we present an example of a quantum phase transition for a Chern insulator that occurs via phase separation and a possible signature of topological edge modes at the phase boundary. Figure 4 shows a series of images near the ($C$, $s$) = (-4, 1) (indicated in Fig. 4A), which has a Kagome lattice structure (Fig. 4B). Upon increasing $V_G$, there is a quantum phase transition between the incompressible Kagome Chern phase to another incompressible phase with petal-like electronic structure [fig. S20, section S13 (*30*)]. Real-space maps show phase separation during this transition. As $V_G$ increases, the Kagome lattice region shrinks while the petal-like region expands (Fig. 4, C to E), ultimately filling the entire field of view (Fig. 4F). The spectroscopy in Fig. 4A shows the closing and opening of a gap when the phase boundary crosses the location of the STM tip (red stars in Fig. 4B-G) as $V_G$ is varied, indicating that the boundary involves a gapless mode as anticipated at the boundary of a Chern insulator(*38*). Although we could not experimentally detect a hysteresis in the gate-dependent spectra for this transition, the observation of phase separation suggests a 1$^{st}$ order phase transition for melting of this interacting Chern phase. Finally, with further electron doping, the petal-like structure evolves into a compressible state exhibiting intricate intra-unit-cell structure [Fig. 4G, fig. S20, section S14 (*30*)].

**Conclusion and outlook**

Our study visualizes a plethora of SBCI states and directly determines the relationship between the quantum indices describing such phases ($C$ and $s$) and the spatial structure of their wavefunctions, showing examples of states that enlarge the unit cell as well as those with complex intra-unit-cell structures. Detailed understanding of what stabilizes these broken symmetry phases awaits further theoretical investigation. We also provide a unique perspective on the quantum phase transitions of correlated and topological phases demonstrating that both Kosterlitz-Thouless-type defect mediated and phase separation-type can occur. Looking ahead, our techniques can provide important new insights when applied to similar experiments on both in-field and zero-field fractional Chern insulators. An interesting setting is the in-field FCI states, as theoretical studies predict topological defects to host fractionalized charges and have non-Abelian statistics(*39*).

**Acknowledgments:**

We thank useful conversations with Taige Wang, Jong Yeon Lee and Michael Zaletel.

**Funding:**

This work was primarily supported by DOE-BES grant DE-FG02-07ER46419, the Gordon and Betty Moore Foundation's EPiQS initiative grants GBMF9469, ARO grant W911NF261A052 to A.Y. Other support for the experimental work was provided by NSF-MRSEC through the Princeton Center for Complex Materials NSF-DMR- 2011750, NSF grant DMR-2312311, NSF grant OMA-2326767ARO MURI (W911NF-21-2-0147), and ONR N00012-21-1-2592. OV was supported in part by the Gordon and Betty Moore Foundation's EPiQS Initiative (Grant No. GBMF11070) and by a grant from the Simons Foundation SFIMPS-NFS00006741-09. A.Y. acknowledge the hospitality of the Trinity College, the Physics Department, and the Cavendish Laboratory in Cambridge U.K. **Author contributions:** MH, Y-CT, AY devised the experiments, with MH, Y-CT created devices structures and carried out the STM measurements and data analysis. RP, and OV carried out the theoretical calculations. KW, and TT provided the h-BN substrates. MH, Y-CT, AY, RP, OV all collaborated in writing of the manuscript. **Competing interests:** Authors declare that they have no competing interests. **Data and materials availability:** The data from this study are available at figshare (*40*).

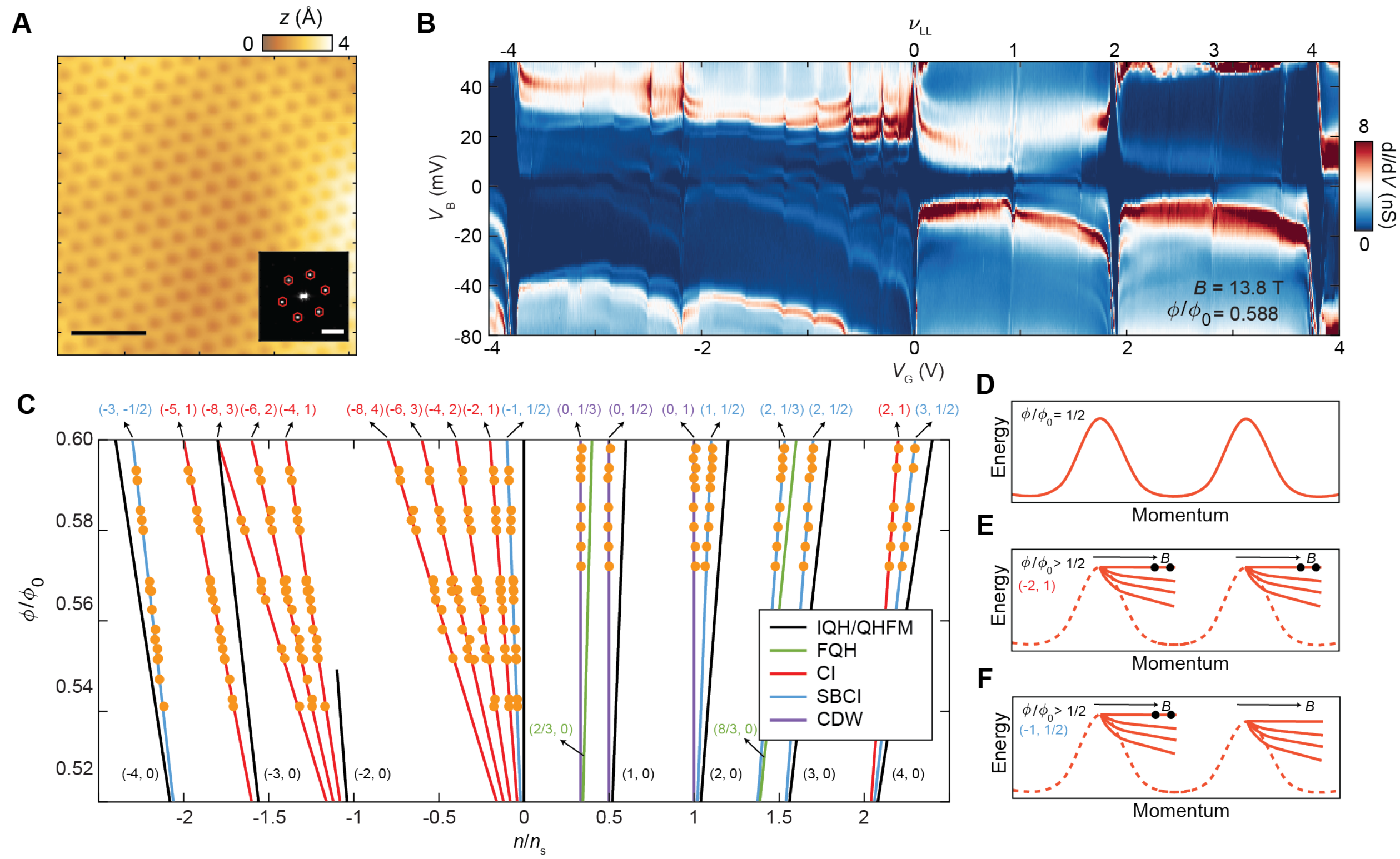

**Fig. 1. Spectroscopic studies of Hofstadter's spectrum in Bernal-stacked bilayer graphene/hBN heterostructure. (A)** A clean topograph of a device with periodic structure of bilayer graphene/hBN moiré. The black scale bar is 50 nm. The inset shows the Fourier transform of the topography map, the Bragg peaks of the moiré are marked by red circles, the corresponding periodicity is 14.28 nm. The white scale bar of the inset is 0.5 nm$^{-1}$. **(B)** A density (tuned by gate voltages $V_G$) dependent STS measured at $B$=13.8 T, $\phi/\phi_o$ = 0.588. The vertical axis is the tip bias voltage $V_B$ where it probes the electron (hole) excitation at $V_B$ >0 ($V_B$ <0). On $V_G$ >0 ($V_G$ <0), the electrons are away from (close to) moiré. The insulating gaps can be identified as gap opening at the Fermi level. Gaps opening at $V_G$ >0 are asymmetric to $V_G$ <0 with respect to the charge neutrality. The top horizontal axis shows the Landau level filling factor $\nu_{LL}$. **(C)** Wannier diagram of the insulating phases observed in density $n/n_s$ dependent STS as a function of $\phi/\phi_o$ (for $\phi/\phi_o$>1/2). The topological indices ($C$, $s$) are listed above and are characterized by formula $n/n_s = C\,\phi/\phi_0 + s$. The colored lines are the corresponding ($C$, $s$) states, and the orange points are the data points. All data are collected at $T_{\mathrm{eff}}$ = 210 mK. **(D)** An illustration of the dispersive band structure at $\phi/\phi_o$=1/2. Two identical magnetic subbands, which are enforced by the magnetic translational symmetry, are shown. **(E)** As the flux increases further for $\phi/\phi_o$>1/2, the minimum of each magnetic subband forms Landau levels. Here, for ($C$, $s$) = (-2, 1) state, the lowest two degenerate Landau levels are fully filled (black dots mean fully filled at field $B$). **(F)** When we half-filled the lowest two degenerate Landau levels, an interaction driven SBCI state ($C$, $s$) = (-1, 1/2) is formed by spontaneously breaking the moiré translational symmetry.

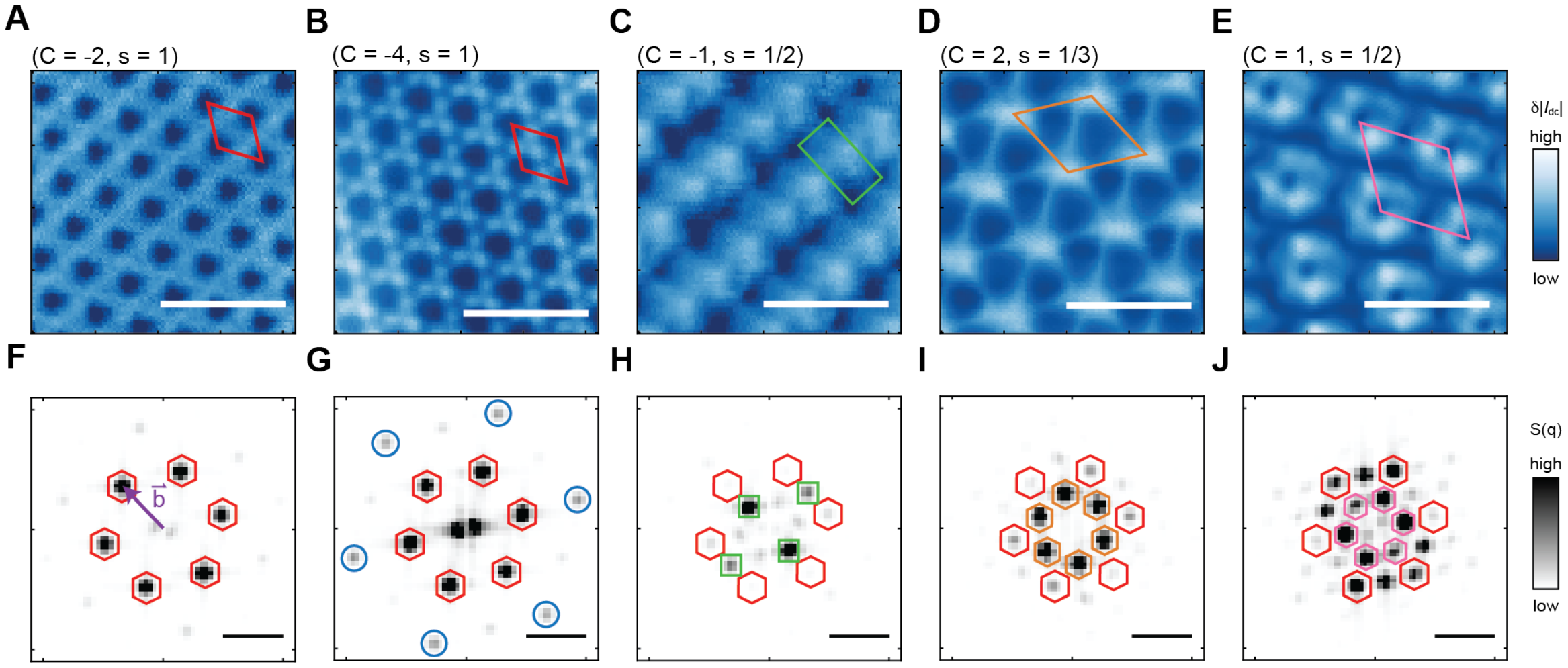

**Fig. 2. Direct imaging of Chern insulators. (A-E)** Real space tunneling current $\delta|I_{\mathrm{dc}}|$ maps measured at **(A-C)** $B$=13.2 T, **(D, E)** $B$=13.8 T and low bias voltages $V_{\mathrm{B}}$ of each ($C$, $s$) states. The white scale bar is 40 nm. **(F-J)** Structure factor $S(\vec{\boldsymbol{q}})$ of the tunneling current modulation $\delta|I_{\mathrm{dc}}|$ in panels A-E, respectively. The scale bar is 0.5 nm$^{-1}$. The bilayer graphene/hBN moiré Bragg peaks are identified by red hexagons. **(A, F)** (-2, 1) state respects the moiré translational symmetry. The red parallelogram is the moiré unit cell. **(B, G)** Emergent Kagome lattice for (-4, 1) state respects the moiré translational symmetry. The Kagome structure shows up as higher harmonics (blue circles) in the $S(\vec{\boldsymbol{q}})$ map. **(C, H)** (-1, 1/2) state the breaks moiré translational symmetry by doubling the unit cell and form a topological stripe phase (green rectangle as the unit cell). The Bragg peaks of the emergent stripe phase are identified as the two brightest green squares. **(D, I)** (2, 1/3) state breaks the moiré translational symmetry by tripling the unit cell and forms a triangular lattice (orange parallelogram as the unit cell). The Bragg peaks are identified as six $\sqrt{3}$ orange hexagons. **(E, J)** (1, 1/2) state the breaks moiré translational symmetry by quadrupling the unit cell and forms a triangular lattice (pink parallelogram as the unit cell). The Bragg peaks of the emergent triangular lattice are identified as the six pink hexagons. All data are collected at $T_{\mathrm{eff}}$ = 210 mK.

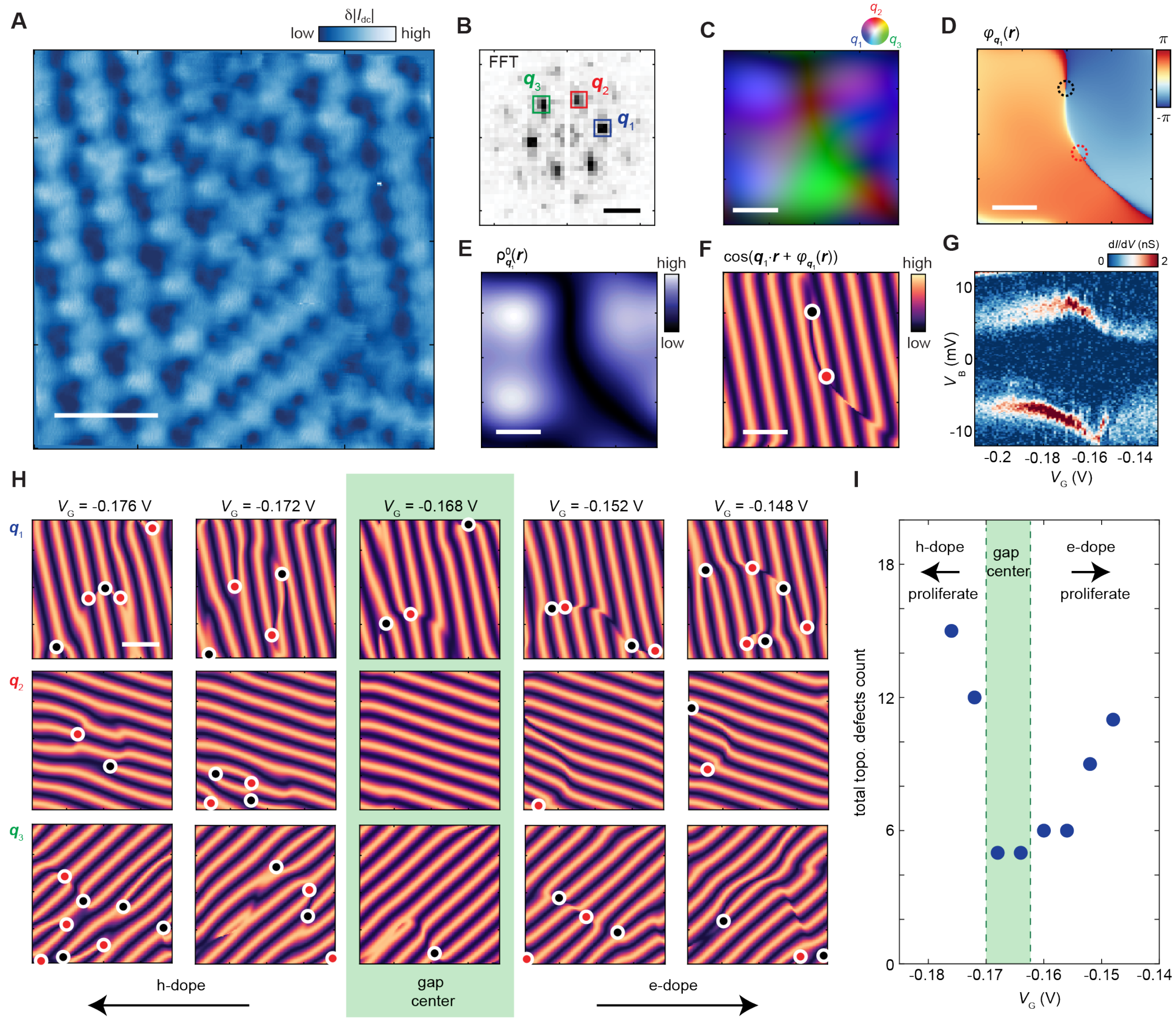

**Fig. 3. Topological melting of a symmetry broken Chern insulator (SBCI) ($C$=-1, $s$=1/2). (A)** A large real space tunneling current $\delta|I_{\text{dc}}|$ map measured at $B$=13.8 T and $V_{\text{B}}$ = 6 mV. The stripes with different orientations can be identified. **(B)** FFT of the $\delta|I_{\text{dc}}|$ map in (A), the stripes with different orientations are identified by $\boldsymbol{q}_{i=1,2,3}$ vectors. The scale bar is 0.25 nm$^{-1}$. **(C)** The real space distribution of normalized amplitude of the three $\boldsymbol{q}_{i=1,2,3}$ vectors. **(D)** The extracted relative phase $\varphi_{\boldsymbol{q}_1}(\boldsymbol{r})$ map. Where the topological defect is identified by dotted black (red) circles, with the clockwise winding of the phase is $2\pi$ ($-2\pi$). The black (red) circle is the dislocation (anti-dislocation). **(E)** The amplitude $\rho^{o}_{\boldsymbol{q}_1}(\boldsymbol{r})$ map of $\boldsymbol{q}_1$. Showing the strength of stripe along $\boldsymbol{q}_1$ direction. **(F)** The $\cos(\boldsymbol{q}_1 \cdot \boldsymbol{r} + \varphi_{\boldsymbol{q}_1}(\boldsymbol{r}))$ map with sharp boundary shifting half of the unit cell. **(G)** A density dependent STS across the insulating gap of SBCI ($C$=-1, $s$=1/2). **(H)** Selected $\cos(\boldsymbol{q}_i \cdot \boldsymbol{r} + \varphi_{\boldsymbol{q}_i}(\boldsymbol{r}))$ maps as a function of $V_{\text{G}}$ across the gap. The top/second/third row is for $\mathbf{q}_1$/ $\mathbf{q}_2$/ $\mathbf{q}_3$ respectively. The dislocations (anti-dislocations) are marked as black (red) dots in the maps. The number of topological defects increases upon doping away from the gap center ($V_{\text{G}}$=-0.168 V). **(I)** The number of topological defects (dislocation + anti-dislocation) as a function of $V_{\text{G}}$. Topological defects proliferate as the system is doped away from the gap center,

suggesting that the SBCI stripe phase melts through a topology-driven mechanism. The white scale bars in all panels are 50 nm. All data are collected at $T_{\mathrm{eff}}$ = 210 mK.

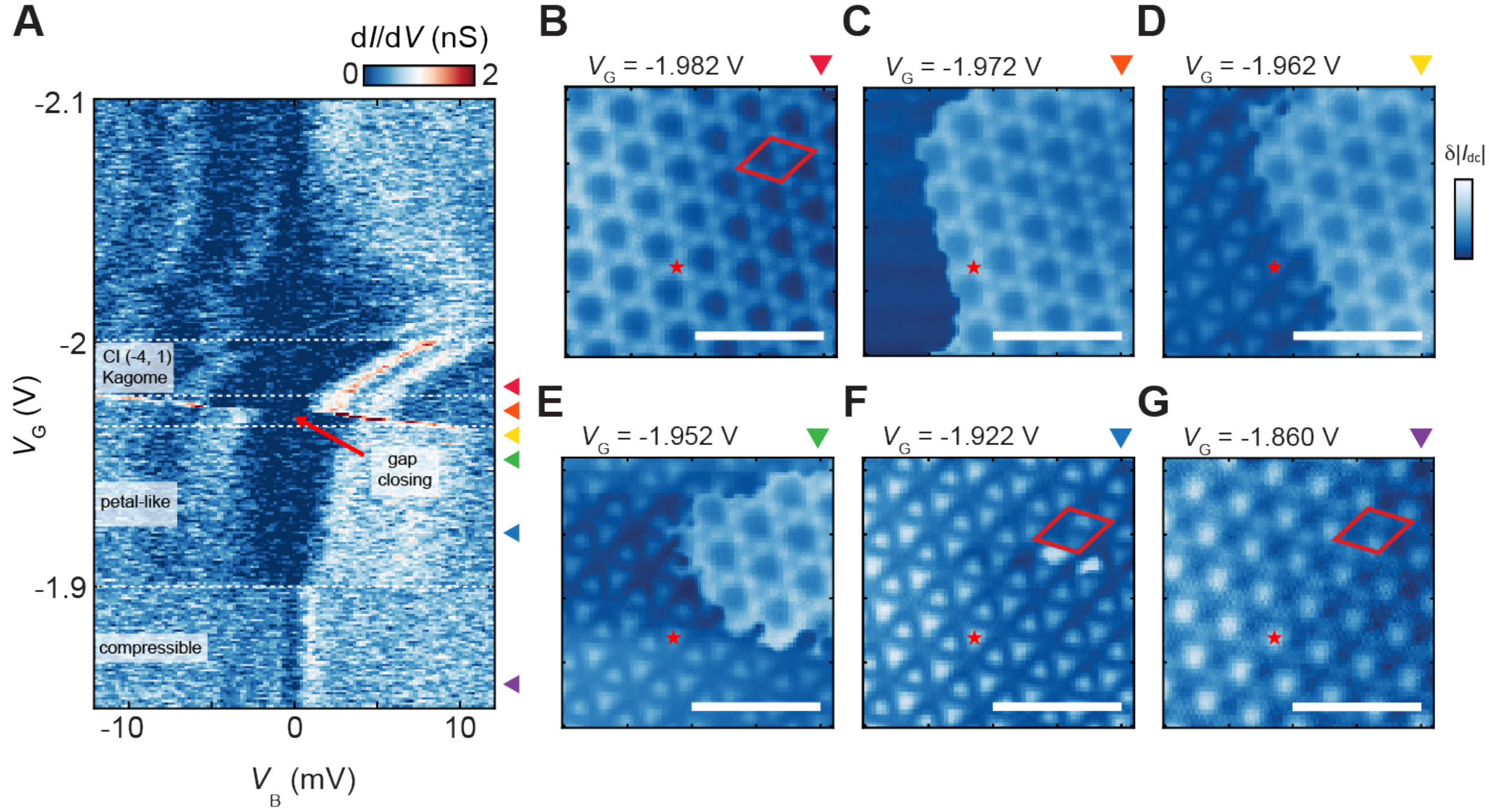

**Fig. 4. Phase separation and quantum melting of a Chern Insulator. (A)** STS near a $(C, s)$ = (-4, 1) state at magnetic field $B$ = 13.2 T. The incompressible spectroscopic features of the Kagome CI as well as gap closing (between orange and yellow arrows) are labeled. The gap closing happens when the sharp phase boundary in (C, D) passes underneath the tip, indicating that the phase boundary is compressible. The gate voltages for images in (B-G) are marked with colored arrows on the right side of the spectrum. **(B)** The tunneling current $\boldsymbol{\delta}|I_{dc}|$ map measured at gate voltage $V_G$= -1.982 V (red) and low bias voltages $V_B$ = 4.8 mV shows an emergent Kagome lattice. The moiré unit cell is marked with red parallelogram. **(C)** The $\boldsymbol{\delta}|I_{dc}|$ map measured at gate voltage $V_G$= -1.972 V (orange) and $V_B$ = 3.6 mV is showing an atomically sharp phase boundary of the Kagome lattice. **(D)** The $\boldsymbol{\delta}|I_{dc}|$ map measured at gate voltage $V_G$= -1.962 V (yellow) and $V_B$ = 5.6 mV is showing a phase boundary between the incompressible Kagome lattice and the incompressible petal-like structure with the Kagome lattice island shrinking. **(E)** The $\boldsymbol{\delta}|I_{dc}|$ map measured at gate voltage $V_G$= -1.952 V (green) and $V_B$ = 5.0 mV. **(F)** The $\boldsymbol{\delta}|I_{dc}|$ map measured at gate voltage $V_G$= -1.922 V (blue) and $V_B$ = 4.0 mV. The entire field of view is filled with petal-like electronic structure. **(G)** The $\boldsymbol{\delta}|I_{dc}|$ map measured at gate voltage $V_G$= -1.860 V (purple) and $V_B$ = 1.6 mV. The system enters a compressible state with complex intra unit cell structure. The scale bars in all $\boldsymbol{\delta}|I_{dc}|$ maps are 40 nm. The red stars in (B-G) indicates the location of the tip while measuring spectrum in (A). All data are collected at $T_{eff}$ = 210 mK.